\documentstyle[12pt]{article}
\def\bbbr{{\rm I\!R}}
\def\bbbc{{\mathchoice {\setbox0=\hbox{$\displaystyle
\rm C$}\hbox{\hbox
to0pt{\kern0.4\wd0\vrule height0.9\ht0\hss}\box0}}
{\setbox0=\hbox{$\textstyle\rm C$}\hbox{\hbox
to0pt{\kern0.4\wd0\vrule height0.9\ht0\hss}\box0}}
{\setbox0=\hbox{$\scriptstyle\rm C$}\hbox{\hbox
to0pt{\kern0.4\wd0\vrule height0.9\ht0\hss}\box0}}
{\setbox0=\hbox{$\scriptscriptstyle\rm C$}\hbox{\hbox
to0pt{\kern0.4\wd0\vrule height0.9\ht0\hss}\box0}}}}
\def\bbbone{{\mathchoice {\rm 1\mskip-4mu l} {\rm 1\mskip-4mu l}
{\rm 1\mskip-4.5mu l} {\rm 1\mskip-5mu l}}}
\renewcommand{\cite}[1]{[\ref{#1}]}
\renewcommand{\bibitem}[1]{\item\label{#1}}
\newcommand{\gB}{g^{\rm B}}\newcommand{\Tr}{{\rm Tr\,}}
\newcommand{\cA}{{\cal A}}
\newcommand{\cB}{{\cal B}}
\newcommand{\cD}{{\cal D}}
\newcommand{\cN}{{\cal N}}
\newcommand{\cT}{{\cal T}}
\newcommand{\cH}{{\cal H}}
\newcommand{\cP}{{\cal P}}
\newcommand{\cS}{{\cal S}}
\newcommand{\re}{\mbox{Re}\,}
\newcommand{\rd}{{\rm d\,\!}}

\newcommand{\bx}{{\bf x}}
\newcommand{\bxt}{{\bf \widetilde{x}}}
\newcommand{\Dt}{{ \widetilde{D}}}
\newcommand{\Lt}{{ \widetilde{L}}}
\newcommand{\Rt}{{ \widetilde{R}}}
\newcommand{\rT}{{\rm T}}
\newcommand{\r}{\varrho}
\newcommand{\gl}[1]{(\ref{#1})}
\newcommand{\nabl}{\nabla \!}
\newcommand{\GaW}{G_\alpha\!W}
\newcommand{\e}[2]{{\rm E}_{#1 #2}}
\newcommand{\rh}[2]{{\rm H}_{#1 #2}}
\newcommand{\rhb}[2]{\overline{{\rm H}}_{#1 #2}}
\begin{document}
\title{Yang-Mills Equation and Bures Metric}
\author{Jochen Dittmann\\Universit\"at Leipzig, Mathematisches Institut}
\date{January 30, 1998}
\maketitle
\begin{abstract}
\noindent
It is shown that the connection form (gauge field)  related to
the generalization of the Berry phase to mixed states proposed by
Uhlmann satisfies the  source-free Yang-Mills equation
$\ast{\rm D}\ast{\rm D \,\omega} =0$, where the Hodge operator is
taken with
respect to the Bures metric on the space of finite dimensional
nondegenerate density matrices.
\end{abstract}
\vspace{0.5cm}
\noindent{\bf Mathematics Subject Classifications (1991):} 53C07, 58E15,
81T13\\
{\bf Key words}:  Yang-Mills equation, Bures metric, purification, state
space, density matrices
\section{Connection form and Bures metric}
In the last years the Riemannian
Bures metric - the quantum analog of the Fisher information in classical
statistics -  became an interesting object of geometrical investigations as
well as of applications. In
this paper we show a further interesting feature of this metric, namely,
that the gauge field defining this metric  on the background of
purifications of mixed states fulfills the source-free Yang-Mills equation.

Let $\cH$ be a Hilbert space of finite dimension $n$. The
 concept of puri\-fi\-cation  of  mixed states  leads in the case of
the algebra  $\cA:=\cB(\cH)$ to the following  principal U($\cH$)-bundle:
The bundle space $\cP^1$ is the manifold of invertible normalized ($\Tr
W^\ast W{=}1$) Hilbert-Schmidt operators
and the base space  $\cD^1$ is the manifold of faithful states on $\cA$.
Define the bundle
projection $\pi:\cP^1\rightarrow
 \cD^1$ by $\pi(W)(a):=\langle W,aW)\rangle_{HS}=\Tr WW^\ast a$, $a\in\cA$.
Actually, by the polar decomposition of the operators $W$ we get a
${\rm U}(\cH)$-principal bundle. $W$ is called a purification of
$\pi(W)$ because it represents a pure state of the algebra
$\cB(\cS_2(\cH))\supset\cA$ and reduces to the state $\pi(W)$ of $\cA$
(\cite{Ulett}). $\cS_2(\cH)$ denotes the space of Hilbert-Schmidt operators.

Although the space of Hilbert-Schmidt operators
coincides with $\cA$ for finite dimension we used this terminology to
emphasize the underlying hermitian product
$\langle\,\cdot\,,\,\cdot\,\rangle_{HS}$
on $\cP^1$. Its  real part  defines  an
${\rm U}(\cH)$-invariant Riemannian metric $g$  on the bundle space $\cP^1$
and gives rise to a  connection on the principal
bundle, declare the horizontal spaces to be  the orthogonal
complements of the vertical directions. The corresponding connection
form (gauge field) we denote by $\omega$. It  was proposed by Uhlmann
generalizing the Berry phase to mixed states (\cite{Ulett}). Its curvature
form ${\rm D}\,\omega$ we denote by $\Omega$. Moreover, the Riemannian
metric and the connection on $\cP^1$ induce a Riemannian metric $\gB $ on
the base space $\cD^1$, define the length of a vector tangent to
$\cD^1$ as the length of any of its horizontal lifts, see formula
\gl{Buresmetrik}. It turns out that this Riemannian metric first given
in \cite{Ugel} is just the Riemannian version of the Bures distance
$\r,\mu\mapsto
d(\r,\mu):=
(2-2\,\Tr\,(\r^\frac{1}{2}\mu\r^\frac{1}{2})^\frac{1}{2})^\frac{1}{2}$
(see \cite{Araki},\cite{Bures}).
More precisely,
$\gB_\r=\frac{1}{2}{\rm Hessian}\{\mu\mapsto
d(\r,\mu)^2\}_{\mu=\r}$ and in local coordinates $\{\mu_i\}$ it is
represented by a half of the  matrix of  second order partial
derivatives of $\mu\mapsto d(\r,\mu)^2$ at $\mu_i=\r_i$.

Finally, note
that the bundle $\pi\,{:}\,\cP^1\rightarrow\cD^1$ is a subbundle of
$\pi\,{:}\,\cP\rightarrow\cD$, where $\cD$ is the manifold of all
faithful positive linear forms on $\cA$ and $\cP$ is the dense in $S_2(\cH)$
subspace of invertible operators.  The connection and the Riemannian
metrics are defined analogously. We  use the same symbols for
the projection, connection, curvature  and metrics on both bundles,
which are by construction Riemannian submersions
(comp.~e.g.~\cite{Besse}).

The aim of this paper is to prove
the following
\vspace{0.2cm}
\newline
{\bf Theorem:} \newline{\sl The curvature form
$\Omega$ on the principal bundle $\cP^1(\cD^1 ,{\rm
U}(\cH),\gB)$
(resp.\ $\cP(\cD,{\rm U}(\cH),\gB)$) fulfills the
source-free Yang-Mills equation
$\ast {\rm D}\ast \Omega=0
$,
where $\ast$ is the horizontal lift of the Hodge star  w.r.~to the Bures
metric $\gB$ on the base space.}
\vspace{0.3cm}
\newline
\noindent
In \cite{DR} it was shown, that the Yang-Mills equation holds for
dim$\cH=2$ in the case of states. The conjecture that this is true for
general $n$ is due to G.~Rudolph. Moreover, for $n=2$ the
Einstein-Yang-Mills equation is fulfilled with a certain cosmological
constant (\cite{TT}). These observations  can be understood from the general
point of view: The connection on $\cP^1$ is reducible to a connection on the
reduced SU($n$)-subbundle $\cP^r:=\left\{W\mid \Tr
WW^\ast=1,\mbox{Det}W>0\right\}\subset\cP^1$  and one verifies that for
$n=2$ (only) the fibers are isometric to SU$(n)$ and totally geodesic (or
in the language of field theory: the scalar fields on $\cD^1$ describing
the vertical part of the bundle metric are constant, compare
e.g.~\cite{Besse}, 9.56/64). Our theorem gives, essentially, a solution
invariant under the natural left U$(n)$-action  of
the YM-equation on the principal bundle ${\rm Gl}(n,\bbbc)\rightarrow
{\rm Gl}(n,\bbbc)/{\rm U}(n)$ (resp.
S$^{2n^2-1}\supset\cP^1\rightarrow\cD^1$), where the base space is
equipped with the  Bures metric. If we regard these solutions as gauge
fields on $\cD^{(1)}$ they are invariant under the induced
U$(n)$-conjugation. By inspection one can see, that the bundles are not
Einstein-Yang-Mills systems for $n>2$.
\section{Notations and preliminary formulae
}First we identify $\cH$ with $\bbbc^n$ and express the above structures
in terms of matrices. The faithful positive linear forms are represented
by nonsingular density matrices;\begin{equation}\cD^{(1)}=\left\{D\in
M_{nn}(\bbbc)\mid D>0 \;(\,\Tr D=1)\;\right\}\,,\end{equation} and the
bundle space becomes \begin{equation}\cP^{(1)}=\left\{W\in
M_{nn}(\bbbc)\mid {\rm Det}W\neq0 \;(\,\Tr
WW^\ast=1)\;\right\}\,.\end{equation} The bundle projection takes
 the form$
\pi(W)=WW^\ast\,
$
and the bundle metric  is
$g(T_1,T_2)=\re\langle T_1,T_2\rangle_{HS}=\frac{1}{2}
(\Tr \,T_1^\ast T_2+T_1 T_2^\ast)$, $ T_i\in\rT\cP$.
The vertical spaces are generated by the U$(n)$-action so that
\begin{equation}
{\rm ker}\,\pi_{\ast W}\quad=\left\{WA\mid A=-A^\ast\in{\rm u}(n)\subset
M_{nn}(\bbbc)\right\}\,.
\end{equation}
For their orthogonal complements one obtains in the not normalized case
\begin{equation}
\rT^{hor}_W\cP:=\left({\rm ker}\,\pi_{\ast W}\right)^\perp=
\left\{GW\mid G=G^\ast\in M_{nn}(\bbbc)\right\}.
\end{equation}
From now on $G$ will be an hermitian and $A$ an antihermitian matrix.
Before we express the further geometric quantities we introduce some
notations.
Let $D:=W W^\ast$ and $\Dt:=W^\ast W$. By $L,R,\Lt,
\Rt$ we denote the operators (depending on $W$) of left
(resp.~right) multiplication by $D$ (resp.~$\Dt$).
Moreover, we use the notations $\bx:=L R^{-1}={\bf Ad}\,D$ and
analogously for $\bxt$. Note that all these operators are
positive, especially the spectrum of $L,\Lt,R,\Rt$ equals the n-fold
spectrum of $D$ whereas the spectrum of $\bx,\bxt$ consists of all quotient
of eigenvalues of $D$. Of course, left and right multiplication operators
commute.

Since the proof of the theorem will be a calculation on the bundle
space we do not need the Bures metric $\gB$ on the base space
explicitly. For completeness we remember
that
\begin{equation}\label{Buresmetrik}\gB=\frac{1}{2}\Tr
\rd\r\frac{1}{L+R}(\rd\r)\,,\quad\mbox{i.~e.~}\quad
\gB(X,Y)=\frac{1}{2}\Tr X G\,,
\end{equation}
where
$
DG+GD=Y$; $ X ,Y\in\rT_D\cD$ (see \cite{Ugel}). But note,
that in affine coordinates (e.g.~using the Pauli matrices for $n=2$)
the metric becomes very complicated for $n>2$ and no good parametrization
seems to be available for general $n$.

$\nabla $
will be the covariant derivative related to the flat metric $g$ on $\cP$.
Later on we will need e.~g.~
\begin{equation}\label{nabla}
\nabl_TW=T\,,\qquad \nabl_TW^\ast=T^\ast\,.
\end{equation}
 Using the Leibniz rule and the parallelity of the matrix
multiplication
 this implies
\begin{equation}
\label{cd}\left(\nabl_{GW}\,\bx\right)(T)=(\nabl_{GW}
D)TD^{-1}-DTD^{-1}(\nabl_{GW}
D)D^{-1}=\left[G+\bx(G),\bx(T)\right]\,.
\end{equation}
Finally, for the covariant derivative $\nabla^1$ of the submanifold $\cP^1$
we have
\begin{equation}\label{nabla1}
\nabl_\cT^{\,1}\cT'= \nabl_\cT\cT'-g(\nabl_\cT\cT',\cN)\cN\, ,
\end{equation}
where $\cN$  is the (normalized) vector field normal to $\cP^1$ given by
$\cN_W=W$.

The connection form of the above described connection equals
\begin{equation}
\omega(T)=\frac{1}{\Lt+\Rt}\left(W^\ast T-T^\ast W\right)\,,\qquad
T\in\rT_W\cP.
\end{equation}
Indeed, if $T$ is vertical, $T=WA$,  then $\omega(T)=A$ and
if $T$ is horizontal, $T=GW$,
then $\omega(T)=0$.
The curvature form takes the following value if its first argument is
horizontal:
\begin{equation}\label{curv}
\!\!\!\!\Omega(GW,T)=2W^\ast\,\frac{1}{1+\bx}
\left(\Big[G,\frac{1}{1+\bx}\Big(T
W^{-1}+\bx\big(W^{-1\ast}T^\ast)\Big)\Big]
\right)\, W^{\ast-1}\,.
\end{equation}
In fact, if $T=WA$ is vertical then $\Omega(GW,WA)=0$. If
$T=G'W$ is horizontal define the horizontal vector fields $\cT$, $\cT'$ by
 $\cT_W=GW$ and $\cT'_W=G'W$. Then
\begin{eqnarray}\label{curv2}
&&\Omega(GW,G'W)=\Omega(\cT,\cT')=
{\rm D}\omega(\cT,\cT')=-\omega([\cT,\cT'])=-\omega(\nabl_{\cT}
\cT'-\nabl_{\cT'} \cT)\nonumber\\&&=\omega([G,G']W)
=2\frac{1}{\Lt+\Rt}\left(W^\ast[G,G']W\right)
=
2W^\ast\,\frac{1}{1+\bx}\left(
\left[G,G'\right]
\right)\, W^{\ast-1}\,
\end{eqnarray}
But the last term equals the right hand side of  \gl{curv} in this
case.
\section{Proof of the theorem}
First we show the assertion in the not
normalized case.
 We have to show $\ast {\rm D}\ast\Omega(T)=0$ for all horizontal vectors
$T$ at $W_o\in\cP$. Let
$T_\alpha:=G_\alpha W_o$, $\alpha=1,\dots, n^2$ be an orthonormal
basis of horizontal vectors at $W_o$. Then
\begin{equation}\label{schritt1}
\ast {\rm D}\ast\Omega(T)=-
\sum_\alpha\left(\nabl_{T_\alpha}\Omega\right)(T_\alpha,T)
\end{equation}and
we will first  deal with the summands on the right hand side. For this
purpose fix $G$ and the $G_\alpha$-s and define  horizontal vector fields
by
 $\cT_{\alpha W}=G_{\alpha}W$, $\cT_ W=GW$.
By
\gl{nabla},\gl{cd}, \gl{curv}, \gl{curv2} and obvious derivation
rules we obtain
\begin{eqnarray}
\Omega(\nabl_{\GaW }\GaW ,GW)&=&
\Omega(G_\alpha^2 W,GW)=
2W^\ast\Big(\frac{1}{1{+}\bx}[G_\alpha^2,G]\Big)W^{\ast-1}\\
\Omega(G_\alpha W ,\nabl_{G_\alpha W}GW)&=&
\Omega(\GaW,GG_\alpha \!W)\nonumber\\&=&
2W^\ast\Big(\frac{1}{1{+}\bx}\Big[G_\alpha,\frac{1}{1{+}\bx}
\big(GG_\alpha+\bx(G_\alpha
G)\big)\Big]\Big)W^{\ast-1}\\
\nabl_{\GaW }\Omega\,(\GaW ,GW)&=&\nabl_{G\alpha W }
\Big(2W^\ast\,\frac{1}{1+\bx}\left(
[G_\alpha,G]\right)\, W^{\ast-1}\Big)\nonumber\\
&=&2W^\ast\Big(\Big[G_\alpha,\frac{1}{1{+}\bx}[G_\alpha,G]\Big]+
\Big(\nabl_{\GaW
}\frac{1}{1{+}\bx}\Big)[G_\alpha,G]\Big)W^{\ast-1}\nonumber\\
&=&2W^\ast
\frac{\bx}{1{+}\bx}\Big(
\Big[G_\alpha,\frac{1{-}\bx}{1{+}\bx}[G_\alpha,G]\Big]\Big)W^{\ast-1}\,.
\end{eqnarray}
Using these formulae we get
\begin{eqnarray}\label{15}
&&\hspace{-1cm}\left(\nabl_{\cT_\alpha}\Omega\right)(\cT_\alpha,\cT)=
(\nabl_{\GaW }\Omega)(\GaW
,GW)\nonumber\\
&&=\nabl_{\GaW }\Omega\,(\GaW ,GW)-
\Omega(\nabl_{\GaW }\GaW ,GW)-\Omega(\GaW ,\nabl_{G_\alpha
W}GW)\nonumber\\
&&=
W^\ast\frac{1}{1+\bx}\Big(
\Big[G_\alpha,\frac{1{-}\bx}{1{+}\bx}[G_\alpha,G]-
\frac{1}{1{+}\bx}\big(GG_\alpha+\bx(G_\alpha G)\big)
\Big]
-[G_\alpha^2,G]
\Big)W^{\ast-1}\,.
\end{eqnarray}
By a suitable choice of the basis of $\cH$ at the very beginning we may
suppose that $W_o=\Lambda U$ with diagonal $\Lambda$ and unitary $U$.
Moreover, by the equivariance of $\ast {\rm D}\ast \Omega$  it is
sufficient to prove the assertion for $U=\bbbone$. Hence, let $W_o$ be
diagonal, say $W_o=\Lambda=\sum_i\lambda_i \e{i}{i}$,
$\lambda_i\in\bbbr_+$, and put
\begin{equation}\label{16}
\rh{i}{j}:=\frac{1}{\sqrt{\lambda_i{+}\lambda_j}}(\e{i}{j}{+}\e{j}{i})
\,,\quad
{\rm H}_i:=\frac{1}{\sqrt{2}}\rh{i}{i}\,,\quad
\rhb{i}{j}:=\frac{\bf
i}{\sqrt{\lambda_i{+}\lambda_j}}(\e{i}{j}{-}\e{j}{i})\,,
\end{equation}
where the $\e{i}{j}$-s are the
 standard  $n\times n$-matrices.
We define the hermitian matrices $G_\alpha$-s as the elements of the set
\begin{equation}\label{H}
\Big\{{\rm H}_i\mid 1\leq i\leq n\}\cup
\Big\{\rh{i}{j}\mid 1\leq i<j \leq n\Big\}\cup
\Big\{\rhb{i}{j}\mid 1\leq i<j \leq n\Big\}\,.
\end{equation}
It is   easy to check that the horizontal vector fields $\cT_\alpha$
are really orthonormal at $\Lambda $. Our next observation
facilitates the summation in
\gl{schritt1}.
\begin{eqnarray}\label{18}
&&\!\!\!\!\!\!\!\!\!\!\!\!
\sum_\alpha\left(\nabl_{T_\alpha}\Omega\right)(T_\alpha,T)\nonumber\\
&=&
\sum_{i}\left(\nabl_{{\rm H}_i W}\Omega\right)({\rm H}_i W,T)
+\sum_{i<j}\left(\nabl_{\rh{i}{j}W}\Omega\right)(\rh{i}{j}W,T)
+\sum_{i<j}\left(\nabl_{\rhb{i}{j}W}\Omega\right)(\rhb{i}{j}W,T)\nonumber\\
&=&\frac{1}{2}\sum_{i,j}
\left(\nabl_{\rh{i}{j}W}\Omega\right)(\rh{i}{j}W,T)+
\left(\nabl_{\rhb{i}{j}W}\Omega\right)(\rhb{i}{j}W,T)\,.
\end{eqnarray}
Inserting \gl{15},\gl{16} into \gl{18} and taking into account that
the operator ${\bf Ad}W^\ast\,(1+\bx)^{-1}$ has a
trivial kernel it remains to show that the sum
\begin{samepage}
\begin{eqnarray}\label{19}
&&\!\!\!\!\!\!\!\!\!\!\!\!\!\!\sum_{i,j,\pm}
\pm\frac{1}{\lambda_i+\lambda_j}\Bigg\{
-[(\e{i}{j}\pm\e{j}{i})^2\,,\,G]\nonumber\\
&&\nonumber\\
&&+\bigg[\e{i}{j}\pm\e{j}{i}\,,\,\frac{1-\bx}{1+\bx}[\e{i}{j}
\pm\e{j}{i}\,,\,G]
-
\frac{1}{1+\bx}\big(G(\e{i}{j}\pm\e{j}{i})+\bx\big((\e{i}{j}\pm\e{j}{i})
G)\big)\bigg]
\Bigg\}
\end{eqnarray}
\end{samepage}
vanishes at the point $\Lambda$ for all hermitean $G$.
Using
$$
\bx(\e{i}{j})_{\,{\textstyle\mid}\,W=\Lambda}=
\frac{\lambda_i}{\lambda_j}\e{i}{j}
$$
and similar formulae for the other operators involving $\bx$ we get after a
straightforward calculation that both partial sums related to the
different signs $\pm$ in \gl{19} vanish individually at $\Lambda$ even for
every $G{=}\e{k}{l}$. This finishes the proof in the not normalized
case.

To proof the assertion in the case of states let $\Lambda\in\cP^1$, $
\sum \lambda_i^2=1$. Of course, formula \gl{schritt1} is valid
independently of the choice of the orthonormal basis
$\left\{T_\alpha\right\}$. Thus we may also set
$G_1=\bbbone$ and complete the list of the $G_\alpha$ suitable. Then
$\cT_{1\,\Lambda}=G_1\Lambda=\Lambda $ is normal to the submanifold
$\cP^1\subset\cP$ and the remaining obtained vectors $G_\alpha
\Lambda$
 are tangent to $\cP^1$ and constitute an
orthonormal basis of horizontal vectors of $\cP^1$ at $\Lambda$. From
\gl{15} we see that the vector $\cT_{1\,\Lambda}{=}\Lambda$
($G_\alpha{=}G_1{=}\bbbone$) does not contribute to \gl{schritt1}.
Moreover, for the remaining vectors $G_\alpha \Lambda$, $\alpha>1$, holds
$\left(\nabl_{G_\alpha\Lambda}\Omega\right)(G_\alpha\Lambda,G\Lambda)=
\left(\nabl_{G_\alpha\Lambda}^{\,1}\Omega\right)(G_\alpha\Lambda,G\Lambda)
$. Indeed, using \gl{nabla1} we obtain
\begin{eqnarray*}
\left(\nabl_{G_\alpha\Lambda}^{\,1}\Omega\right)
(G_\alpha\Lambda,G\Lambda)&=&
\left(\nabl_{G_\alpha\Lambda}\Omega\right)(G_\alpha\Lambda,G\Lambda)\\
&-&g(\nabl_{G_\alpha\Lambda}G_\alpha\Lambda,\Lambda)\,\Omega
(G\Lambda,\Lambda)
+g(\nabl_{G_\alpha\Lambda}G\Lambda,\Lambda)\,\Omega(G_\alpha
\Lambda,\Lambda)\,.
\end{eqnarray*}
But the last two additional terms vanish by \gl{curv2} (set $G'=\bbbone$
in \gl{curv2}). Thus the
 vanishing of \gl{schritt1} in the not normalized case implies the
vanishing in the case of states. This finishes the proof of the
theorem.
\newline
I would like to thank G. Rudolph and A. Uhlmann for helpful discussions.
\\\\
\noindent{\Large\bf References}
\setlength{\labelwidth}{7cm}

{\bf Address:}\\
J.~Dittmann,\\
Mathematisches Institut, Universit\"at Leipzig\\
Augustusplatz 10/11, 04109 Leipzig\\
e-mail: {\tt dittmann@mathematik.uni-leipzig.de}

\begin{enumerate}
\bibitem{Araki} Araki, H.: Bures distance
function and a generalization of Sakai's noncommutative Radon-Nikodym
theorem, Publ.~Res.~Inst.~Math.~Sci.~S Kyoto Univ.~{\bf 8} (1972),
335-362.
\bibitem{Besse}{Besse, A.L.: }{\sl Einstein Manifolds},
Springer-Verlag, Berlin, Heidelberg, 1987.
\bibitem{Bures}Bures, D.: An extension of Kakutani's theorem on
infiniteproduct measures to the tensor product of semifinite
$w^*$-algebras, Trans.~Amer.~Math.~Soc.~{\bf 135} (1969), 199-212.
\bibitem{D1}Dittmann, J.: Some properties of the Riemannian
Bures metric on mixed states, J.~Geom.~Phys.~{\bf 13} (1994), 203-206.
\bibitem{DR}Dittmann, J., Rudolph, G.: On a connection governing
parallel transport along $2\times 2$-density matrices,
J.~Geom.~Phys.~{\bf 10} (1992), 93-106.
\bibitem{Eguchi}Eguchi, T., Gilkey, P., Hanson, A.: Gravitation,
Gauge Theories and Differential Geometry, Phys.~Rep.~ {\bf 66} (1980),
213-393.
\bibitem{TT}Rudolph, G., Tok, T.:  A
Certain Class of Einstein-Yang-Mills Systems, Rep.~Math. Phys.~ {\bf 39}
(1997), 433-446.
\bibitem{Kob}Kobayashi, S.,Nomizu, K.: {\sl Foundations of Differential
Geometry, Vol.I/II}, Interscience Publisher, New York, London, 1963.
\bibitem{Trautmann}Trautmann, A.: Fibre bundles, gauge fields, and
gravitation, Gen.~Rel.~Grav.~{\bf 1 }(1990), 287-308.
\bibitem{Ulett}Uhlmann, A.:  A Gauge Field
Governing Parallel Transport Along Mixed States, Lett.~Math.~Phys.~{\bf
21} (1991), 229-236.
\bibitem{Ugel}Uhlmann, A.: The metric of Bures and the Geometric Phase,
in: Gielerak, R. et al.~(eds.): {\sl Groups and Related Topics}, Kluwer
Academic Publishers, 1992, 267-274
\end{enumerate}
\end{document}